# EVOLUTION OF SPECTRA FOR MECHANICAL AND WIND WAVES IN A LARGE TANK

**Vladislav Polnikov**[1,†,1], **Fangli Qiao**[2,3], **Hongyu Ma**[2], **and Shumin Jiang**[2]

[1]A.M. Obukhov Institute of Atmospheric Physics of RAS, Moscow, Russia
[2]First Institute of Oceanography of MNR, Qingdao, China
[3]Laboratory for Regional Oceanography and Numerical Modeling, Qingdao National Laboratory for Marine Science and Technology, Qingdao, China

Empirical spectra for mechanical and wind waves measured in a large tank of the First Institute of Oceanography of China are presented. Analysis for the first and the second type of waves is done separately. It is shown that, in the case of mechanical waves with a steepness more than 0.2, the frequency spectra of waves evolve to ones with the tail decay $S(f) \propto f^{-4.2 \pm 0.1}$, whilst the shape of spectra at large fetches is self-similar. Numerical solutions of the four-wave kinetic equation, written in the fetch-limited version, result in the same spectra. This allows treating the empirical observations for mechanical waves as the natural evolution of free nonlinear surface waves. In the case of wind waves, the wave spectra evolve to ones with the tail decay $S(f) \propto f^{-4.0 \pm 0.05}$ at fetches $X$ greater than 8 meters, under any applied winds. The well-known Toba's "three-second relation", $H \propto T_p^{3/2}$, between the mean wave height, $H$, and the peak period, $T_p$, is well fulfilled. Though, the intensities of spectra tails do not follow the Toba's ratio, $S(f) \propto (gu_*)f^{-4}$, rather they better correspond to the ratio $S(f) \propto (u_*^2/Xg)^p (gu_*) f^{-4}$ with $p = 1/3$. Moreover, the shape of the wind-wave spectra is not self-similar. Discussion of the results is presented.

## 1. Introduction

Laboratory tanks are widely used for studying properties of both mechanical and wind waves, and numerous processes in the air-sea boundary layer (see references in Donelan et al, 2004; Babanin and Haus 2009; Longo et al. 2012; Mitsuasu 2015; Zavadsky & Shemer 2017, among others). The real advantage of tank studies resides in a good control of the ambient state. Due to this preference, the tank measurements allow to find numerous fine results dealing with the following phenomena: drift currents (Wu, 1975); features of the wind-wave evolution (Toba 1972, 1973; Zavadsky at al. 2017; Zavadsky & Shemer 2017); air-sea interaction phenomena (Donelan et al. 2004; Troizkaya et al. 2012); wind and current profiles, and turbulence parameters at the air-sea interface (Longo 2012; Longo et al. 2012); dissipation rate of the turbulence kinetic energy in a water layer under wavy surface (Babanin & Haus 2009; Polnikov et al. 2019); and wave-induced vertical mixing (Dai et al. 2010). These empirical results gave rise to numerous theoretical investigations aimed to explain observations (e.g., Kitagorodskii 1983; Phillips 1985; Zakharov 2017; Polnikov 2019; among numerous others). But not all of empirical results are clear till now, and they are needed of further investigation, confirmation, and elaboration.

To this aim, we have executed our own laboratory experiment with mechanical and wind waves, devoted to studying a wide scope of problems, including: wave- and wind-induced drift currents; vertical profiles of currents in their relation to the turbulence production under a wavy surface; and empirical estimations of the turbulent diffusivity provided by the wave motions. Unexpectedly, it turns out that even initial data processing, related to the wave spectrum evolution, revealed a lot of interesting results, some part of which is closely related to ones in (Toba 1973).

---

[1]† Correspondent author is Vladislav Polnikov: polnikov@mail.ru

For mechanical waves, we found that on a length of the tank, under certain initial conditions, the wave frequency spectra, $S(f)$, evolve to have the tail decay $S(f) \propto f^{-4.2 \pm 0.1}$. Herewith, for the initial wave steepness greater 0.2, the spectral shape becomes self-similar at fetches $X$ greater 12 meters. This empirical effect is revealed for the first time. Below, it is treated as the explicit manifestation of the four-wave nonlinear interactions for surface waves.

In the case of wind waves, the wave spectra evolve to ones with the tail decay $S(f) \propto f^{-4.0 \pm 0.05}$. This effect is realized at fetches $X$ greater 8 meters for any applied wind speeds. The well-known "three-second relation", $H \propto T_p^{3/2}$ (Toba, 1972), between the mean wave height, $H$, and the peak period, $T_p$, is well fulfilled. Though, the intensities of spectra tails do not follow the Toba's ratio, $S(f) \propto (gu_*) f^{-4}$ (Toba, 1973), for measured values of the friction velocity, $u_*$. The spectra tails correspond better to the ratio $S(f) \propto (u_*^2/Xg)^p (gu_*) f^{-4}$, with the power parameter $p = 1/3$. It means the stronger dependence of the spectral intensity on $u_*$ at fixed fetch $X$, with respect to the mentioned Toba's dependence. Additionally, owing to a large size of the tank used, the fetch-dependence of the wind-wave spectral tail intensity is also revealed. Moreover, we found that the shape of the wind-wave spectra, $S(f)$, is not self-similar.

As far as all these results have reasonable physical importance, we present them here in a short version with details sufficient for the express information.

Layout of the paper is the following. In Sect. 2, the equipment and details of measurements are described. In Sect. 3, results for mechanical waves are presented, followed by Sect. 4 with the proper analysis. Sect. 5 is devoted to the wind-wave spectra results, the analysis of which is given in Sect. 6. A short discussion of the results is presented in Sect. 7. Conclusions are given in the Abstract.

## 2. Equipment and measurements description

In our experiment, we have used the wind-wave tank of the First Institute of Oceanography of MNR (FIO), located in Qingdao, China. A sketch of the tank is shown in figure 1.

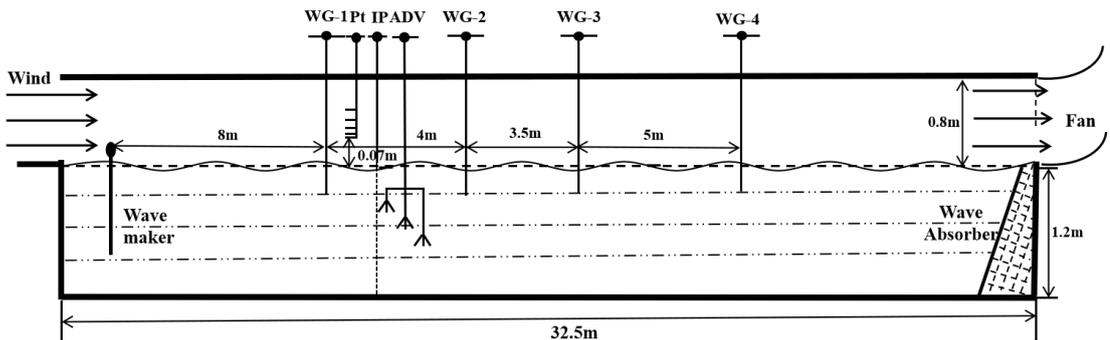

FIGURE 1. A sketch of the tank. WGs are the wave gauges, PT are the Pitot tubes, ADV is the set of the Acoustic Doppler Velocimeters, IP is the set for injecting in water float-particles or color dye.
The WGs locations and sizes of the tank, water depth, and air space are shown. The width of the tank is 1 m.

For the wave measuring, the capacity wave–gauges (WG) were used. The Pitot tubes (PT) and three Acoustic Doppler Velocimeters (ADV) were used for measuring the wind- and current-profiles, respectively. Locations of WGs, sizes of the tank, water depth, and air space are shown in figure 1. Locations of WG1-WG4 are below marked as points P1-P4, respectively.

The wind is provided by the fan located at the outlet of the tank. Closed circulation of the wind was realized (not shown). PT have five channels located at horizons z1 =7 cm, z2 = 10cm, z3 = 14 cm, z4 = 25 cm, and z5 = 40cm. For wind greater 10 m/s, all tubes were 1.2 cm upper.

Two kinds of mechanical waves were studied: regular, quasi monochromatic waves (MR), and waves with a wide spectral band (MW) alike Pierson-Moscowiz (PM) or JONSWAP (JW) wave spectra (Komen et al, 1994). These waves were provided by the wave maker with three initial frequencies: 1.5, 1.0, and 0.7 Hz, and with 3-5 amplitudes at each initial frequency: from $H_S$ = 1 cm to $H_S$ =20 cm (here $H_S$ is the wave-maker equivalent of a significant wave height for generated waves). All these wave parameters were provided by the wave-maker computer.

Another type of waves was the wind-waves (WW) generated by the fan wind, $W$. The values of $W$ are approximately equal to the wind speed at the mid of the air part of the tank (z = 40 cm). Five values of wind $W$ were applied: 4, 6, 8, 10, and 12 m/s.

The WG-records were 10 minutes long, with the sampling of 50 Hz. The same duration was used for measurements with PT, which have the record-averaging of 1 min. Processing the WG and PT data were done within the MATLAB software. For the wave-spectra evaluation, the AR- and the Welch-method were used. As the data-series have 30 thousands points (more than a thousand of dominant periods), the 95% confidence intervals are very small. In the bi-logarithmic coordinates, they are [+10%, -12%] and [+15%, -20%] for the AR-method and the Welch-method, respectively. According to the statistical theory, the proper std-errors for the spectral intensities are about 3% and 5%. As well known (Kay, 1988), both methods are equivalent for estimating the spectral tail shape, though the AR-method is preferable for the peak-parameters estimating, whilst the Welch-method does for the spectrum-tail intensity.

| Fetch | 8 m | | | | 12 m | | | | 20.5 m | | | |
|---|---|---|---|---|---|---|---|---|---|---|---|---|
| | | | | | Puddle frequency $f_p$ =1.5 Hz | | | | | | | |
| Puddle $H_S$, cm | a, cm | $f_p$, Hz | $k_p$, r/m | $ak_p$, n/d | a, cm | $f_p$, Hz | $k_p$, r/m | $ak_p$, n/d | a, cm | $f_p$, Hz | $k_p$, r/m | $ak_p$, n/d |
| 3 | 1.41 | 1.49 | 8.9 | 0.126 | 1.36 | 1.49 | 8.9 | 0.12 | 1.15 | 1.49 | 8.9 | 0.10 |
| 5 | 2.31 | 1.49 | 8.9 | 0.21 | 2.13 | 1.49 | 8.9 | 0.19 | 1.93 | 1.48 | 8.8 | 0.17 |
| 7 (30%) | 2.95 | 1.48 | 8.9 | 0.26 | 2.76 | 1.48 | 8.9 | 0.25 | 2.45 | 1.43 | 8.2 | 0.20 |
| 10 (50%) | 3.46 | 1.48 | 8.9 | 0.31 | 3.20 | 1.48 | 8.8 | 0.28 | 2.44 | 1.36 | 7.4 | 0.18 |
| | | | | | Puddle frequency $f_p$=1.0Hz | | | | | | | |
| 3 | 1.09 | 1.0 | 4.0 | 0.044 | 1.03 | 1.0 | 4.0 | 0.04 | 0.98 | 1.0 | 4.0 | 0.04 |
| 5 | 1.85 | 1.0 | 4.0 | 0.074 | 1.74 | 1.0 | 4.0 | 0.07 | 1.66 | 1.0 | 4.0 | 0.07 |
| 7 | 2.60 | 1.0 | 4.0 | 0.105 | 2.45 | 1.0 | 4.0 | 0.10 | 2.37 | 1.0 | 4.0 | 0.09 |
| 10 | 3.63 | 1.0 | 4.0 | 0.15 | 3.48 | 1.0 | 4.0 | 0.14 | 3.39 | 1.0 | 4.0 | 0.14 |
| 15 | 5.21 | 1.0 | 4.0 | 0.21 | 5.03 | 1.0 | 4.0 | 0.20 | 4.94 | 0.99 | 4.0 | 0.20 |
| | | | | | Puddle frequency $f_p$=0.7Hz | | | | | | | |
| 15 | 3.70 | 0.7 | 1.97 | 0.073 | 3.70 | 0.7 | 1.97 | 0.07 | 3.63 | 0.7 | 1.97 | 0.07 |
| 20 | 4.80 | 0.7 | 1.97 | 0.095 | 5.03 | 0.7 | 1.97 | 0.10 | 4.80 | 0.7 | 1.97 | 0.09 |

TABLE1. Main parameters for mechanical regular waves

Note. In the first column, the percentage of breaking is shown in brackets. The steepness values are shadowed.

### 3. Results for mechanical waves

#### 3a. Regular waves (MR-waves)

The main parameters for MR-waves are given in table 1, where $a = \left(2\int S(f)df\right)^{1/2}$ is the mean wave amplitude, $f_p$ is the peak frequency, $k_p$ is the peak wave-number, and $ak_p$ is the mean steepness.

## Evolution of Spectra

As one can see, the initial steepness $ak_p$ for MR-waves varies widely, having a reasonable decrease if the starting values are greater 0.2. For these values of $ak_p$, some visible plunging breaking takes place, though this makes no remarkable impact on the wave-spectrum evolution.

The main feature of the wave-spectrum evolution is the achieving spectral shape with the tail decay $S(f) \propto f^{-4.2\pm0.1}$ at the tank fetches (even at point P3 = 15.5 m) for the runs with initial $ak_p$ greater 0.2. The typical frequency spectra, $S(f)$, for the runs with the puddle parameters $Hs = 3-10$ cm and $f_p = 1.5$ Hz are shown in figure 2(a). After normalization $S(f)$ by $S(f_p)$ and $f$ by $f_p$, the spectral shape becomes selfsimilar, $S_{sf}(f)$ (figure 2(b)).

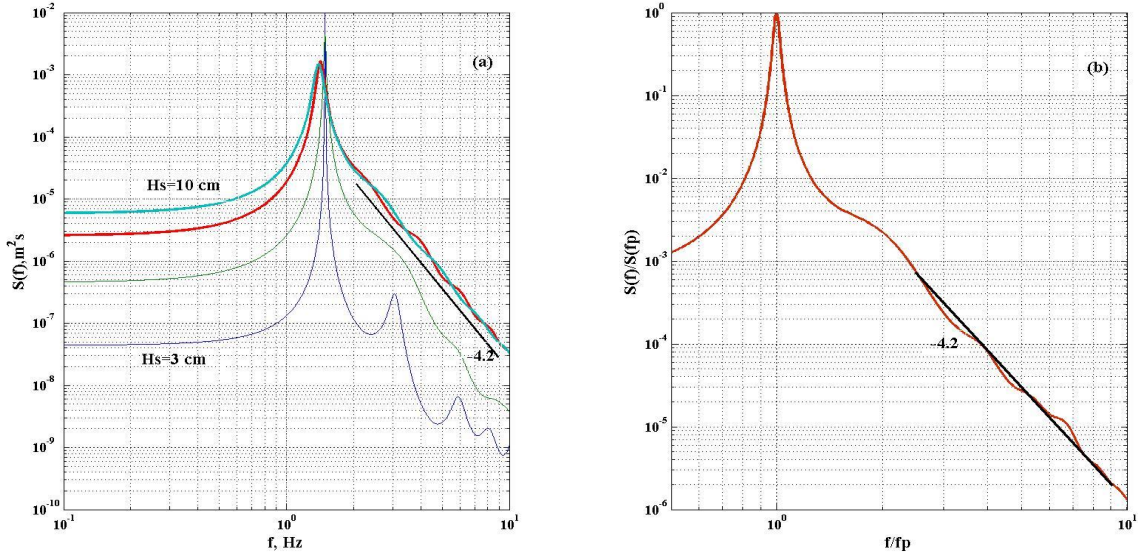

FIGURE 2. (a) MR-waves spectra at P4 = 20.5 m for runs with the puddle $Hs = 3 - 10$ cm and initial $f_p = 1.5$ Hz. (b) normalized MR-waves spectra at P4 for runs $Hs = 7$ and 10 cm, and initial $f_p = 1.5$ Hz. Bold black lines with digits show the law of the power-decay for spectra tails (the error is about 2-3%).

For waves with the same $Hs$ but $f_p \leq 1.0$ Hz, having initial $ak_p < 0.2$, the form $S_{sf}(f)$ is not achieved at point P4 = 20.5 m, possibly, due to a rather small fetch for these waves in the tank.

### 3b. Wide band spectra waves (MW-waves)

Parameters for MW-waves are very similar to ones shown in table 1, for this reason they are not presented here. The main difference is very small initial steepness $ak_p$ (for PM-spectra they are of the order 0.05). Herewith, in the case of JW-spectrum with $Hs = 15$ cm and $f_p = 1.0$ Hz, $ak_p$ becomes of the order of 0.11. It is only in this case, the spectral shape for MW-waves evolved radically, having a tendency to reach the self-similar form. This evolution is shown in figure 3, where it is seen that for the fetches longer the length of the tank, the self-similar shape $S_{sf}(f)$ has a tendency to be eventually established.

## 4. Analysis of results for mechanical waves

Establishing the self-similar spectral shape resembles the known results of numerical solutions of the four-wave kinetic equation (KE) (e.g., Badulin et al, 2005; Polnikov and Qiao, 2019). Thus, this effect can be treated as the typical evolution of the nonlinear surface waves.

To check this idea, we have solved numerically the fetch-limited version of the KE, which, in the stationary case, has the kind

$$(g/2\omega)\partial S(\omega)/\partial x = I_{NL}(S), \qquad (1)$$

where $I_{NL}(S)$ is the four-wave kinetic integral (see the last references). It was found that for a proper initial steepness and very narrow angular distribution proportional to $\cos^{20}(\theta-\theta_0)$, indeed, numerical spectra achieve the self-similar shapes at the fetches of the order 20 m and more. Though, the numerical frequency width of the spectral peak is twice greater (figure 4).

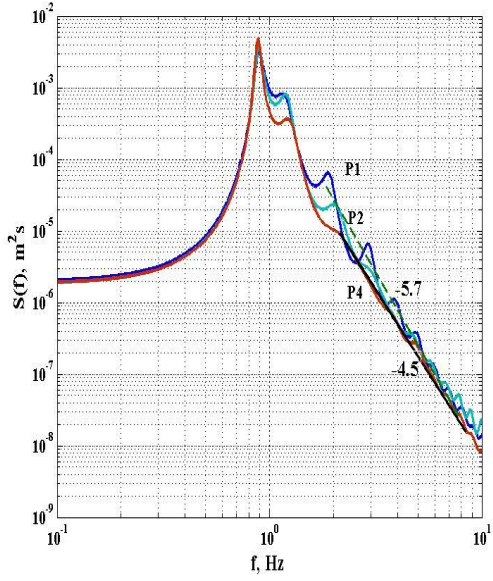 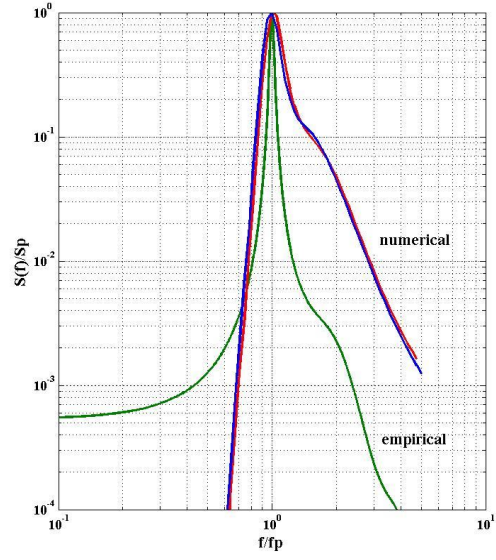

FIGURE 3. The spectrum shapes at points P1, P2, and P4, for the case of JW-spectrum for MW-waves with $H$s = 15 cm and $f_p$ = 1.0 Hz

FIGURE 4. Normalized empirical and numerical spectra. Two coinciding lines for the latter means the normalized numerical spectra at fetches of 15 and 20 m for MR-waves with $H$s =7 cm and $f_p$ =1.5 Hz.

This difference is not so important from the physical viewpoint and becomes clear, if one take into attention a discrepancy between the empirical and numerical initial spectral shapes, which is stipulated by the numerical restrictions of using the KE (Zakharov, 2017). Despite this discrepancy, in our mind, numerical results allow us to conclude that the observed empirical effect (formatting the self-similar spectrum shape for mechanical waves in the tank) is the real evidence of natural evolution of the nonlinear surface waves. This statement finalizes the current analysis for mechanical waves, the extended discussion of which will be continued in Sect. 7.

## 5. Results for wind waves

The wind waves were generated by the fan for five different speeds, *W*: 4, 6, 8, 10, and 12 m/s. The main parameters of wind waves are shown in table 2 (intensity of breaking was not fixed).

| Position | P1 = 8 m | | | | P2 = 12 m | | | | P4 = 20.5 m | | | |
|---|---|---|---|---|---|---|---|---|---|---|---|---|
| Fan wind *W*, m/s | a, cm | $f_p$, Hz | $k_p$, r/m | $ak_p$, n/d | a, cm | $f_p$, Hz | $k_p$, r/m | $ak_p$, n/d | a, cm | $f_p$, Hz | $k_p$, r/m | $ak_p$, n/d |
| 4 | 0.29 | 4.84 | 94.2 | 0.27 | 0.53 | 3.58 | 51.5 | 0.27 | 0.75 | 2.56 | 26.3 | 0.20 |
| 6 | 0.50 | 3.82 | 58.7 | 0.29 | 0.77 | 2.89 | 33.6 | 0.26 | 1.06 | 2.17 | 18.9 | 0.20 |
| 8 | 0.74 | 3.36 | 45.4 | 0.34 | 1.01 | 2.64 | 28.0 | 0.28 | 1.53 | 2.0 | 16.1 | 0.25 |
| 10 | 1.01 | 2.85 | 32.6 | 0.33 | 1.33 | 2.32 | 21.6 | 0.29 | 2.09 | 1.78 | 12.7 | 0.26 |
| 12 | 1.24 | 2.74 | 30.2 | 0.37 | 1.65 | 2.06 | 17.0 | 0.28 | 2.54 | 1.67 | 11.2 | 0.28 |

TABLE 2. Main parameters for wind waves

Note. The steepness values are shadowed.

## Evolution of Spectra

The proper growth relations for the dimensionless energy, $\tilde{E} = (a^2/2)g^2/W^4$, and the spectrum peak frequency, $\tilde{\omega}_p = (2\pi f_p)W/g$, with the dimensionless wave fetch, $\tilde{X} = Xg/W^2$, have a kind typical for wind waves (Komen et al, 1994, Toba, 1973). In terms of fan wind $W$, they are

$$\tilde{E} = 7.5 \cdot 10^{-3}\tilde{X}^{1.05\pm0.1} \quad \text{and} \quad \tilde{\omega}_p = 16.0\tilde{X}^{-0.35\pm0.3}, \qquad (2)$$

what correlates very well with the Toba's ratio, $H_* = 6.2 \cdot 10^{-2} T_{*p}^{3/2}$, if one uses in (2) the normalization by $u_*$ with the drag coefficient $C_d = 10^{-3}$. Thus, for the growth laws, we have no new results. Though, results for the spectral shapes have some novelty.

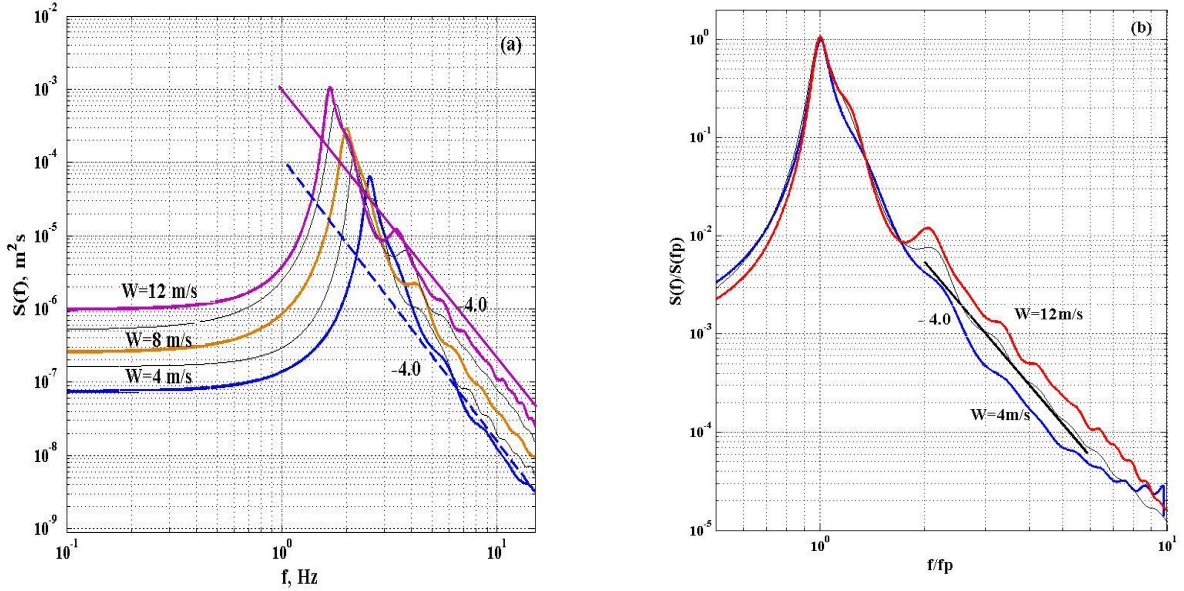

FIGURE 5. The ensemble of wind-wave spectra at point P4 =20.5 m. (a) ordinary spectra; (b) nomalized spectra. Bold black lines with digits show the law of the power-decay for spectra tails (the error is about 2-3%).

First of all, consider the ensemble of spectra for winds $W$ = 4-12 m/s at final point P4 (figure 5(a)). One can see that all spectra have the same decay law, $S(f) \propto f^{-4.0\pm0.05}$, for $f > 2f_p$. Though, in this range, a very wide splitting of intensities takes place. Herewith, the "-4" slopes, shown by the inclined strait lines in figure 5(a), do not cover the peak range of the spectra. This result is the same as in figure 4 in (Toba, 1973), though, it seems that the spectral shape is not self-similar. This fact is evidently seen in figure 5(b) for the normalized spectra (only three spectra are shown for simplicity). This is the first radical deviation of our results from the statement in Toba(1973) that the wind waves in a tank have a self-similar shape.

Second deviation we found after multiplying the spectra by the factor $f^4$ (figures 6(a,b)). To interpret them, we attracted estimations of the friction velocities $u_*$ realized in our tank. They are shown in table 3 (errors are about 20% for $W \leq 6$ m/s and 10% otherwise). Note that in different part of the tank, the values of $u_*$ differ nearly in 1.5 times for wind $W$ = 4 m/s, and do in 2 times for $W$ = 12 m/s. (Similar results were presented, with no analysis, in Longo (2012), figure 4). These facts are very important for treating the spectra tails results..

From figures 6(a, b), it is clearly seen the following: (i) the intensity of the auxiliary value, $I = S(f) \cdot f^4$, depends on $u_*$ stronger than linearly: nearly five times in figure 6(a) and 10 times in figure 6(b); (ii) intensity $I$ depends on fetch $X$ (explicitly seen for $W$ = 4 and 8 m/s in figure 6(b)). These are the new results not described by Toba(1973).

## 6. Analysis of results for wind waves

Treating the results for wind waves is based on the estimations of $u_*$. From table 3, it is seen that this value strongly changes along the tank what is typical for the tank experiments (Toba, 1972; Longo, 2012). Using the constancy of value $I = S(f)f^4$ and better estimations of $u_*$, compared with Toba(1972) (it seems that $u_*$ presented in table 1 of (Toba, 1972) are overestimated), allow us to quantify the stronger dependence $I$ on $u_*$, shown in figures 6(a,b).

| Fan wind, m/s | Location PT at X = 9 m | | Location PT at X= 20 m | |
|---|---|---|---|---|
| | Mean $u_*$, cm/s | Mean $W_{10}$, m/c | Mean $u_*$, cm/s | Mean $W_{10}$, m/c |
| 4 | 8.2 | 4.8 | 12.5 | 5.3 |
| 6 | 11.7 | 6.6 | 17.5 | 7.6 |
| 8 | 14.4 | 8.5 | 27.8 | 10.3 |
| 10 | 20.4 | 10.8 | 40 | 13.3 |
| 12 | 28.2 | 12.5 | 53 | 15.8 |

TABLE 3. Estimations for $u_*$ and $W_{10}$ in the tank

Herewith, a small visible deviation of the upmost tail in figure 6(b) from a fixed value can be induced by the Doppler shift of frequency *f* due to wind drift. As noted in (Donelan et al, 1985), this effect plays a minor role for waves with *f* less 10 Hz, therefore it is neglected hereafter.

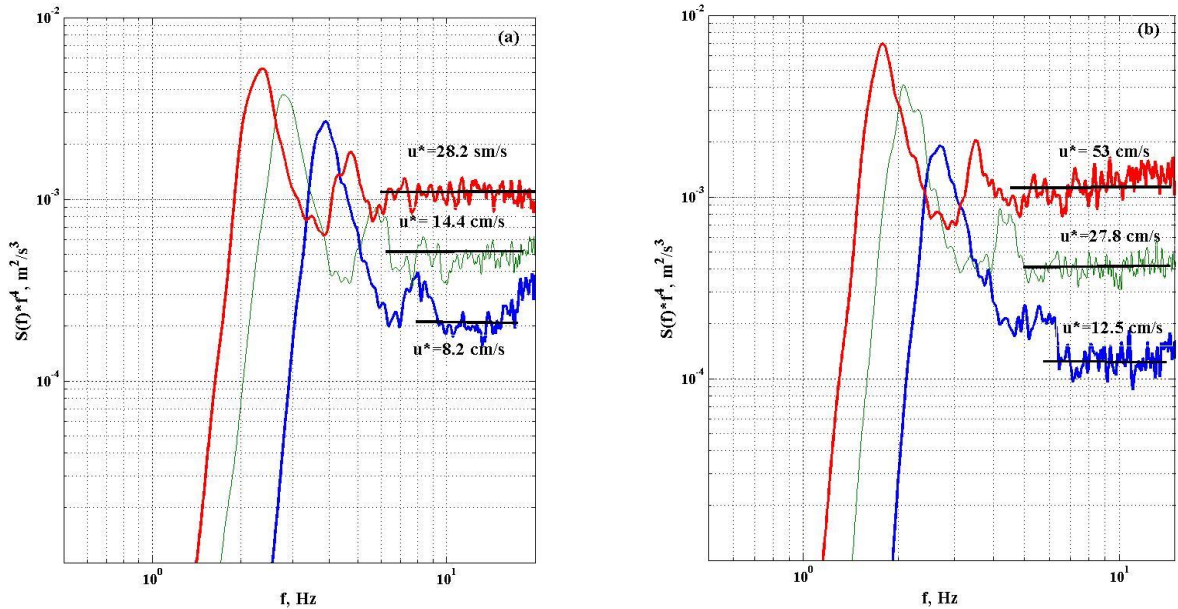

FIGURE 6. Wind-wave Welch-spectra multiplied by the factor $f^4$ : (a) point P2, and (b) point P4.

The quantified tail intensities of our spectra $S(f,u_*)$ could be written in the Toba's form
$$S(f,u_*) = C_T g u_* f^{-4}, \qquad (3)$$
with $C_T$ varying in range $(1 \div 3) \times 10^{-4}$. As seen from figures 6(a,b), form (3) does not accurately corresponds to our results for any fixed $C_T$. To treat this fact, we supposed that the Toba's form

(3) should be generalized to form $S(f,u_*) = F(f_p,u_*)gu_* f^{-4}$, with the dimensionless factor $F(f_p,u_*)$ that does not change the tail dependence, $S(f) \propto f^{-4}$. The simplest version, $F(f,u_*) = C_n(f_p u_*/g)^n$, with the parameter $n = 1$, results in the tail of the form

$$S(f,u_*) = C_1(f_p u_*/g)gu_* f^{-4} = C_1(u_*^2/Xg)^{1/3}gu_* f^{-4} \quad , \qquad (4)$$

where the second ratio of (2) is used. The form (4) with fixed value $C_4 = 4.1 \times 10^{-3}$ corresponds to the results shown in figures 6(a,b) much better (errors are less than 50%) than form (3) (errors are more 200%). Moreover, the form (4) gives the generalization of form (3) to the fetch-dependence for the wind-wave spectra tail, the presence of which was widely discussed earlier (e.g., Donelan et al, 1985).

The said above finalizes treating the wind-fetch-dependence for the wind-wave spectra tail.

The self-similarity absence for the spectral shape needs no long discussion, as it is evidently seen from figure 5(b). Some extent of self-similarity takes place in the spectral peak range, only. That allows us to state that the full wind-wave spectra in a tank cannot be found as the self-similar solution of pure KE(1), in contrary to statements in (Badulin et al, 2005; Zakharov, 2017).

## 7. Discussion

First, we continue the analysis of origin the self-similar spectra for mechanical waves. Despite of some extent of wave breaking and inevitable destruction the wave-motion potentiality, in our mind, the executed numerical simulations allows considering mechanical waves in a tank as free ones. Results of the KE (1) solutions, shown in figure 4, support this viewpoint.

Note that in the case of quasi-unidirected mechanical waves, the traditional KE should result in the zero evolution, due to vanishing the four-wave kinetic integral kernel having the exact resonances for both four frequencies and wave-vectors (Chalikov, 2012; Zakharov, 2017). But in reality, the exact resonance between four waves involved in the interaction, as shown in (Zaslavskii and Polnikov, 1998), is not necessary, and small deviations from the resonance may lead to a non-zero energy transfer and wave evolution. This effect was numerically confirmed in (Chalikov, 2012), what closes the physical part of the empirical results treatment. Detailed numerical manifestation of this effect is out of this paper, as it needs a separate consideration.

Regarding the wind waves, we should add the following. For long time, the Toba's conclusion about the self-similarity of the wind-wave spectra shape in a tank with the spectral tail (3) was widely recognized. This fact was treated as the Kolmogorov spectra (Kitagorodskii, 1983; Phillips, 1985) following from the KE solution (Badulin et al, 2005; Zakharov, 2017). But recently, it was numerically proved (Polnikov, 2018) that the wind-wave spectra tails could have very different forms, as the tail slopes are fully defined by the input *In* and dissipation *Dis* terms added to term $I_{NL}(S)$ in the r.h.s of KE (1). Formally, the spectral tail (4) can be numerically simulated if proper parameterizations for the *In* and *Dis* terms are chosen. Though, the physical reason of robust formation the spectral tail $S(f) \propto f^{-4}$ in a tank, for any fetches and winds, is not clear. This makes a great challenge for theoreticians.

The answer to this challenge could reside in accounting the following two facts. First, the dimension of value $I = S(f)f^4$ is [L$^2$/T$^3$], what corresponds to the dimension of dissipating rate of the turbulence kinetic energy, $\varepsilon$. Thus, the constancy of $I$ means a constancy of $\varepsilon$ in the tail range of wind-wave spectrum. Second, the value $I$ is proportional to $S(f)f^4 df$, i.e., to the second power of the local steepness, $a^2(f)k^2(f)$. This means that the value of wave steepness is constant in the tail range. Both these facts could prompt the solution of the posed problem.

Finally, we note that the absence of self-similarity for the wind-wave spectra in a tank, shown in figure 5(b), and the significant generalization of the Toba's result (3) up to result (4), followed from figures 6(a,b), of course, are needed their confirmation and elaboration in future experiments. The results for mechanical waves are needed their detailed numerical simulations.

*Acknowledgments*. The authors thank students Wang Hue and Li Chao for their assistance in the experiment execution. The work was done under support of the Russian Foundation for Basic Research, grant # 18-05-00161, and NSFC-Shandong Joint Fund grant of China, # U1606405.

REFERENCES

Babanin, A.V. & Haus B.K. 2009 On the existence of water turbulence induced by non-breaking surface waves. *J. Phys. Oceanogr.* 2009. **39. P.** 2675–2679. doi:10.1175/2009JPO4202.

Badulin, S.I., Pushkarev, A.N., Resio, D., & Zakharov, V.E. 2005 Self-similarity of wind-driven seas. *Nonlinear Process. Geophys.* **12**, 891-945. SRef-ID: 1607-7946/npg/2005-12-891

Chalikov, D, 2012 On the nonlinear energy transfer in the unidirected adiabatic surface waves. *Phys. Lett.* **376**, 2795–2816. https://doi.org/10.1016/j.physleta.2012.08.026.

Dai, D., Qiao, F., Sulisz, W., Han L., & Babanin, A. An Experiment on the Nonbreaking Surface-Wave-Induced Vertical Mixing. *J. Phys. Oceanogr.* 2010. **40**, 2180-2188.

Donelan, M. A., Haus, B. K., Reul, N., Plant, W. J., Stiassnie, M., Graber, H. C., Brown, O. B. & Saltzman, E. S. 2004 On the limiting aerodynamic roughness of the ocean in very strong winds. *Geophys. Res. Lett.,* **31**, L18306. doi:10.1029/2004GL019460.

Donelan, M.A, Hamilton, J., & Hui W.H. 1985 Directional Spectra of Wind-Generated Waves. *Phil. Trans. R. Soc. London.* **A315**, 509-562

Kay, S. M. 1988. *Modern Spectral Estimation, Theory and Application*, 543 p. Englewood Cliffs, NJ: Prentice Hall, New Jersey.

Kitaigorodski, S. A. 1983 On the theory of the equilibrium range in the spectrum of wind-generated gravity waves. J. Phys. Oceanogr. **13** (5), 816–827.

Komen G.I., Cavaleri L., Donelan M., Hasselmann K., Hasselmann S., & Janssen P.A.E.M. 1994 *Dynamics and Modelling of Ocean Waves*, 554 p. Cambridge University Press,.

Longo, S. 2012 Wind-generated water waves in a wind tunnel: Free surface statistics, wind friction and mean air flow properties. *Coast. Eng.* **61**, 27-41. doi:10.1016/j.coastaleng.2011.11.008.

Longo, S., Chiapponi, L., Clavero, M., Mäkel, T., Liang, D. 2012 Study of the turbulence over the air-side and water-side boundary layers in experimental laboratory wind induced surface waves. *Coast. Eng.* **69**, 67-81. doi:10.1016/j.coastaleng.2012.05.012

Mitsuasu, H. 2015 Reminiscences on the study of wind waves Proc. *Jpn. Acad., Ser. B*, **91** , 109-130. doi: 10.2183/pjab.91.109.

Phillips, O. M. 1985 Spectral and statistical properties of the equilibrium range in wind-generated gravity waves. *J. Fluid Mech.* **156,** 505–531.

Polnikov, V.G. 2018 The Role of Evolution Mechanisms in the Formation of a Wind-Wave Equilibrium Spectrum *Izvestiya, Atmospheric and Oceanic Physics,* **54(**4), 394–403. doi**:** 10.1134/S0001433818040278.

Polnikov V.G. 2019 A Semi-Phenomenological Model for Wind-induced Drift Currents. *Boundary-Layer Meteorol*. 172(3), 417-433. https://doi.org/10.1007/s10546-019-00456-1.

Polnikov V.G., Baidakov, G.A. & Troitskaya, Yu. I. 2019 Dissipation Rate of Turbulence in a Water Layer under Wind Waves Based on Data of a Laboratory Experiment. *Izvestiya, Atmospheric and Oceanic Physics,* **55**(5), 492–501. doi: 10.1134/S0001433819050104.

Polnikov, V.G. & Qiao, F. 2019 The kinetic equation solutions and Kolmogorov spectra. *IOP Conf. Ser.: Earth Environ. Sci.* **231,** 012043. doi:10.1088/1755-1315/231/1/012043.

Toba, Y. 1972 Local balance in the air–sea boundary processes. I: On the growth process of wind waves. *J. Oceanogr. Soc. Jpn*. **28**, 109–121.

Toba, Y. 1973 Local balance in the air–sea boundary processes. III: On the spectrum of wind waves. *J. Oceanogr. Soc. Jpn*. **29**, 209–220.

Troitskaya, Yu.I., Sergeev, D.A., Kandaurov, A.A., Baidakov, G.A., Vdovin, M.A. & Kazakov, V.I. 2012 Laboratory and theoretical modeling of air–sea momentum transfer under severe wind conditions. *J. Geophys. Res*. 117, C00J21. doi:10.1029/2011JC007778.

Wu, J. 1975 Wind-induced drift currents. *J. Fluid Mech*. **68,** 49-70.

Zakharov, V.E. 2017 Analytic theory of wind–driven sea. *Science direct. Procedia UITAM*. Available on-line at: www. sciencedirect.com (www.elsevier. com/locate/ procedia).

Zaslavskii, M.M. & Polnikov,V.G. 1998. Three-Wave Quasi-Kinetic Approximation in the Problem of the Evolution of a Spectrum of Nonlinear Gravity Waves at Small Depths. *Izvestiya, Atmospheric and Oceanic Physics,* **34**(5), 609–616.

Zavadsky, A., Benetazzo, A. & Shemer, L. 2017 On the two-dimensional structure of short gravity waves in a wind wave tank. *Phys. Fluids* 29 (1), 016601.

Zavadsky, A. & Shemer, L. 2017 Water waves excited by near-impulsive wind forcing. *J Fluid Mech*. **828**, 459-495. doi:10.1017/jfm.2017.521